\documentclass[number,seceqn,citesort,dvips]{arxstspdf}
\usepackage{mathrsfs}
\usepackage{graphics,flushend,stfloats}
\volume{25}
\issue{1}
\pubyear{2010}
\firstpage{72}
\lastpage{87}
\doi{10.1214/10-STS322}

\makeatletter
\newtheorem{theorem}{Theorem}
\newtheorem{proposition}{Proposition}

\newproclaim{example}{Example}
\newproclaim{remark}{Remark}
\newproclaim{definition}{Definition}
\newcommand{\Prob}{\mathsf{P}}
\newcommand{\bel}{\operatorname{Bel}}
\newcommand{\pl}{\operatorname{Pl}}

\newcommand{\Unif}{\operatorname{Unif}}
\newcommand{\Bet}{\operatorname{Beta}}

\newcommand{\Exp}{\operatorname{Exp}}

\newcommand{\A}{\mathcal{A}}
\newcommand{\BB}{\mathscr{B}}

\newcommand{\R}{\mathbb{R}}
\newcommand{\ZZ}{\mathbb{Z}}

\newcommand{\E}{\mathcal{E}}
\newcommand{\M}{\mathscr{M}}
\newcommand{\X}{\mathbb{X}}
\newcommand{\U}{\mathbb{U}}
\renewcommand{\S}{\mathcal{S}}
\renewcommand{\SS}{\mathscr{S}}
\newcommand{\TT}{\mathbb{T}}

\newcommand{\PP}{\mathbb{P}}

\newcommand{\FF}{\mathbb{F}}

\newcommand{\Utilde}{\widetilde{U}}

\newcommand{\Qbar}{\overline{Q}}
\newcommand{\phibar}{\overline{\varphi}}

\renewcommand{\phi}{\varphi}

\newcommand{\eqref}[1]{(\ref{#1})}

\makeatother

\begin{document}
\begin{frontmatter}

\title{Dempster--Shafer Theory and Statistical Inference with Weak Beliefs}
\runtitle{Statistical Inference with Weak Beliefs}

\begin{aug}
\author[a]{\fnms{Ryan} \snm{Martin}\ead[label=e2]{rgmartin@math.iupui.edu}},
\author[b]{\fnms{Jianchun} \snm{Zhang}\ead[label=e3]{zhang10@stat.purdue.edu}}
\and
\author[c]{\fnms{Chuanhai} \snm{Liu}\ead[label=e1]{chuanhai@stat.purdue.edu}\corref{}}

\address[a]{Ryan Martin is Assistant Professor, Department of Mathematical
Sciences, Indiana University--Purdue University Indianapolis, 402 North
Blackford Street, Indianapolis, Indiana 46202, USA \printead{e2}.}
\address[b]{Jianchun Zhang is Ph.D. Candidate, Department of Statistics,
Purdue University, 250 North University Street, West Lafayette, Indiana
47907, USA \printead{e3}.}
\address[c]{Chuanhai Liu is Professor, Department of Statistics, Purdue
University, 250 North University Street, West Lafayette, Indiana 47907, USA
\printead{e1}.}

\runauthor{R. Martin, J. Zhang and C. Liu}

\affiliation{Indiana University--Purdue University Indianapolis, Purdue
University and Purdue University}

\end{aug}

%
\begin{abstract}
The Dempster--Shafer (DS) theory is a powerful tool for probabilistic
reasoning based on a formal calculus for combining evidence. DS theory
has been widely used in computer science and engineering applications,
but has yet to reach the statistical mainstream, perhaps because the DS
belief functions do not satisfy long-run frequency properties.
Recently, two of the authors proposed an extension of DS, called the
weak belief (WB) approach, that can incorporate desirable frequency
properties into the DS framework by systematically enlarging the focal
elements. The present paper reviews and extends this WB approach. We
present a general description of WB in the context of inferential
models, its interplay with the DS calculus, and the maximal belief
solution. New applications of the WB method in two high-dimensional
hypothesis testing problems are given. Simulations show that the WB
procedures, suitably calibrated, perform well compared to popular
classical methods. Most importantly, the WB approach combines the
probabilistic reasoning of DS with the desirable frequency properties
of classical statistics.
\end{abstract}


\begin{keyword}
\kwd{Bayesian}
\kwd{belief functions}
\kwd{fiducial argument}
\kwd{frequentist}
\kwd{hypothesis testing}
\kwd{inferential model}
\kwd{nonparametrics}.
\end{keyword}

\end{frontmatter}
%

\section{Introduction}
\label{Sintro}

A statistical analysis often begins with an iterative process of
model-building, an attempt to understand the observed data. The end
result is what we call a \textit{sampling model}---a model that describes
the data-generating mechanism---that depends on a set of unknown
parameters. More formally, let $X \in\X$ denote the observable data,
and $\Theta\in\TT$ the parameter of interest. Suppose the sampling
model $X \sim\Prob_\Theta$ can be represented by a pair consisting of
(i) an equation
%
\begin{equation}
\label{eqaeqn}
X = a(\Theta, U),
\end{equation}
where $U \in\U$ is called the \textit{auxiliary variable}, and (ii)~a
probability measure $\mu$ defined on measurable subsets of $\U$. We
call \eqref{eqaeqn} the \textit{a-equation}, and $\mu$ the \textit
{pivotal measure}. This representation is similar to that of
Fraser \cite{fraser1968}, and familiar in the context of random data
generation, where a random draw $U \sim\mu$ is mapped, via \eqref
{eqaeqn}, to a variable $X$ with the prescribed distribution
depending on known $\Theta$. For example, to generate a random variable
$X$ having an exponential distribution with fixed rate $\Theta= \theta
$, one might draw $U \sim\Unif(0,1)$ and set $X = -\theta^{-1} \log
U$. For inference, uncertainty about $\Theta$ is typically derived
directly from the sampling model, without any additional
considerations. But Fisher \cite{fisher1935b} highlighted the
fundamental difference between sampling and inference, suggesting that
the two problems should be, somehow, kept separate. Here we take a new
approach in which inference is not determined by the sampling model
alone---a so-called \textit{inferential model} is built to handle
posterior uncertainty separately.

Since the early 1900s, statisticians have strived for inferential
methods capable of producing posterior probability-based conclusions
with limited or no prior assumptions. In Section \ref{Sfiducialds} we
describe two major steps in this direction. The first major step,
coming in the 1930s, was Fisher's fiducial argument, which uses a
``pivotal quantity'' to produce a posterior distribution with no prior
assumptions on the parameter of interest. Limitations and
inconsistencies of the fiducial argument have kept it from becoming
widely accepted. A~second major step, made by Dempster in the 1960s,
extended both Bayesian and fiducial inference. Dempster uses \eqref
{eqaeqn} to construct a probability model on a~class of subsets of
$\X\times\TT$ such that conditioning on $\Theta$ produces the
sampling model, and conditioning on the observed data $X$ generates a
set of upper and lower posterior probabilities for the unknown
parameter $\Theta$. Dempster \cite{dempster2008} argues that this
uncertainty surrounding the exact posterior probability is not an
inconvenience but, rather, an essential component of the analysis. In
the 1970s, Shafer \cite{shafer1976} extended Dempster's calculus of
upper and lower probabilities into a general theory of evidence. Since
then, the resulting Dempster--Shafer (DS) theory has been widely used
in computer science and engineering applications but has yet to make a
substantial impact in statistics. One possible explanation for this
slow acceptance is the fact that the DS upper and lower probabilities
are \textit{personal} and do not satisfy the familiar long-run frequency
properties under repeated sampling.

Zhang and Liu \cite{zl2010} have recently proposed a variation of DS
inference that does have some of the desired frequency properties. The
goal of the present paper is to review and extend the work of Zhang and
Liu~\cite{zl2010} on the theory of statistical inference with \textit
{weak beliefs} (WBs). The WB method starts with a belief function on
$\X\times\TT$, but before conditioning on the observed data~$X$, a
weakening step is taken whereby the focal elements are sufficiently
enlarged so that some desirable frequency properties are realized. The
belief function is weakened only enough to achieve the desired
properties. This is accomplished by choosing a ``most efficient''
belief function from those which are sufficiently weak---this belief is
called the \textit{maximal belief} (MB) solution.

To emphasize the main objective of WB, namely, modifying belief
functions to obtain desirable frequency properties, we present a new
concept here called an \textit{inferential model} (IM). Simply put, an IM
is a belief function that is bounded from above by the conventional DS
posterior belief function. For the special case considered here, where
the sampling model can be described by the a-equation \eqref{eqaeqn}
and the pivotal measure~$\mu$, we consider IMs generated by using
random sets to predict the unobserved value of the auxiliary variable~$U$.

The remainder of the paper is organized as follows. Since WBs are built
upon the DS framework, the necessary DS notation and concepts will be
introduced in Section \ref{Sfiducialds}. Then, in Section \ref
{SgeneralWB}, we describe the new approach to prior-free posterior
inference based on the idea of IMs. Zhang and Liu's WB method is used
to construct an IM, completely within the belief function framework,
and the desirable frequency properties of the resulting MB solution
follow immediately from this construction. Sections \ref{Stesting} and
\ref{Snonparametrics} give detailed WB analyses of two
high-dimensional hypothesis testing problems, and compare the MB
procedures in simulations to popular frequentists methods. Some
concluding remarks are made in Section \ref{Sdiscuss}.

\section{Fiducial and Dempster--Shafer Inference}
\label{Sfiducialds}

The goal of this section is to present the notation and concepts from
DS theory that will be needed in the sequel. It is instructive, as well
as of historical interest, however, to first discuss Fisher's fiducial argument.

\subsection{Fiducial Inference}
\label{SSfiducial}

Consider the model described by the a-equation \eqref{eqaeqn}, where
$\Theta$ is the parameter of interest, $X$ is a sufficient statistic
rather than the observed data, and $U$ is the auxiliary variable,
referred to as a pivotal quantity in the fiducial context. A crucial
assumption underlying the fiducial argument is that each one of
$(X,\Theta,U)$ is uniquely determined by \eqref{eqaeqn} given the
other two. The pivotal quantity $U$ is assumed to have an a priori distribution $\mu$, independent of $\Theta$. Prior to the
experiment, $X$ has a sampling distribution that depends on $\Theta$;
after the experiment, however, $X$ is no longer a random variable. To
produce a posterior distribution for $\Theta$, the variability in $X$
prior to the experiment must somehow be transferred, after the
experiment, to $\Theta$. As in Dempster \cite{dempster1963}, we
``continue to believe'' that $U$ is distributed according to $\mu$
after $X$ is observed. This produces a distribution for $\Theta$,
called the fiducial distribution.

\begin{example}
\label{exnormalmean}
To see the fiducial argument in action, consider the problem of
estimating the unknown mean of a $N(\Theta,1)$ population based on a
single observation $X$. In this case, we may write the a-equation \eqref
{eqaeqn} as
\[
X = \Theta+ \Phi^{-1}(U),
\]
where $\Phi(\cdot)$ is the cumulative distribution function (CDF) of
the $N(0,1)$ distribution, and the pivotal quantity $U$ has a priori distribution $\mu= \Unif(0,1)$. Then, for a fixed $\theta$,
the fiducial probability of $\{\Theta\leq\theta\}$ is, as Fisher \cite
{fisher1930} reasoned, determined by the following logical sequence:
\begin{eqnarray*}
\Theta\leq\theta\quad& \iff&\quad  X-\Phi^{-1}(U) \leq\theta\\
& \iff&\quad  U \geq\Phi
(X-\theta).
\end{eqnarray*}
That is, since the events $\{\Theta\leq\theta\}$ and $\{U \geq\Phi
(X-\theta)\}$ are equivalent, their probabilities must be the same;
thus, the fiducial probability of $\{\Theta\leq\theta\}$, as
determined by ``continuing to believe,'' is $\Phi(\theta-X)$. We can,
therefore, conclude that the fiducial distribution of $\Theta$, given~$X$, is
%
\begin{equation}
\Theta\sim N(X,1).
\label{eqnormalfiducialposterior}
\end{equation}
Note that \eqref{eqnormalfiducialposterior} is exactly the objective
Bayes answer when $\Theta$ has the Jeffreys (flat) prior. A more
general result along these lines is given by Lindley \cite{lindley1958}.
\end{example}

For a detailed account of the development of\break Fisher's fiducial
argument, criticisms of it, and a comprehensive list of references, see
Zabell \cite{zabell1992}. For more recent developments in fiducial
inference, see Hannig \cite{hannig2009}.

\subsection{Dempster--Shafer Inference}
\label{SSds}

The Dempster--Shafer theory is both a successor of Fisher's fiducial
inference and a generalization of Bayesian inference. The foundations
of DS have been laid out by Dempster \cite
{dempster1966,dempster1967,dempster1968a,dempster2008} and Shafer \cite
{shafer1976,shafer1978,shafer1979,shafer1981,shafer1982}. The DS theory
has been influential in many scientific areas, such as computer science
and engineering. In particular, DS has played a major role in the
theoretical and practical development of artificial intelligence. The
2008 volume \textit{Classic Works on the Dempster--Shafer Theory of Belief
Functions} \cite{yagerliu2008}, edited by R. Yager and L. Liu, contains
a selection of nearly 30 influential papers on DS theory and
applications. For some recent statistical applications of DS theory,
see Denoeux \cite{denoeux2006}, Kohlas and Monney \cite
{kohlasmonney2008} and Edlefsen, Liu and Dempster \cite{edlefsen2009}.

DS inference, like Bayes, is designed to make probabilistic statements
about $\Theta$, but it does so in a very different way. The DS
posterior distribution is not a probability distribution on the
parameter space $\TT$ in the usual (Bayesian) sense, but a distribution
on a collection of subsets of $\TT$. The important point is that a
specification of an a priori distribution for $\Theta$ is
altogether avoided---the DS posterior comes from an a priori
distribution over this collection of subsets of $\X\times\TT$ and the
DS calculus for combining evidence and conditioning on observed data.

Recall the a-equation \eqref{eqaeqn} where $X \in\X$ is the observed
data, $\Theta\in\TT$ is the parameter of interest, and $U \in\U$ is
the auxiliary variable. In this setup, $X$, $\Theta$ and $U$ are
allowed to be vectors or even functions; the nonparametric problem
where the parameter of interest is a CDF is discussed in Section \ref
{Snonparametrics}. Here $X$ is the full observed data and not
necessarily a reduction to a sufficient statistic as in the fiducial
context. Furthermore, unlike fiducial, the sets
%
\begin{eqnarray}
\label{eqsets}
\TT_{x,u} & =& \{\theta\in\TT\dvtx x=a(\theta,u)\},
\nonumber
\\[-8pt]
\\[-8pt]
\nonumber
\U_{x,\theta} & =& \{u \in\U\dvtx x=a(\theta,u)\}
\end{eqnarray}
are not required to be singletons.

Following Shafer \cite{shafer1976}, the key elements of the DS analysis
are the \textit{frame of discernment} and \textit{belief function};
Dempster \cite{dempster2008} calls these the \textit{state space model}
and the \textit{DS model}, respectively. The frame of discernment is $\X
\times\TT$, the space of all possible pairs $(X,\Theta)$ of real-world
quantities. The belief function $\bel\dvtx 2^{\X\times\TT} \to[0,1]$ is
a set-function that assigns numerical values to events $\E\subset\X
\times\TT$, meant to represent the ``degree of belief'' in $\E$.
Belief functions are generalizations of probability measures---see
Shafer \cite{shafer1976} for a full axiomatic development---and
Shafer \cite{shafer1979} shows that one can conveniently construct
belief functions out of suitable measures and set-valued mappings
through a ``push-forward'' operation. For our statistical inference
problem, a particular construction comes to mind, which we now describe.

Consider the set-valued mapping $M\dvtx \U\to2^{\X\times\TT}$ given by
%
\begin{equation}
\label{eqfocal}
\quad M(U) = \{(X,\Theta) \in\X\times\TT\dvtx X = a(\Theta,U)\}.
\end{equation}
The set $M(U)$ is called a \textit{focal element}, and contains all those
data-parameter pairs $(X,\Theta)$ consistent with the model and
particular choice of $U$. Let $\M= \{M(U)\dvtx U \in\U\} \subseteq2^{\X
\times\TT}$ denote the collection of all such focal elements. Then the
mapping $M(\cdot)$ in \eqref{eqfocal} and the pivotal measure $\mu$ on
$\U$ together specify a belief function
%
\begin{equation}
\label{eqbel}
\quad\bel(\E) = \mu\{U\dvtx  M(U) \subseteq\E\},\quad   \E\subset\X\times\TT.
\end{equation}
Some important properties of belief functions will be described below.
Here we point out that $\bel$ in \eqref{eqbel} is the push-forward
measure $\mu M^{-1}$, and this defines a probability distribution over
measurable subsets of $\M$. Therefore, when $U \sim\mu$, one can think
of $M(U)$ as a \textit{random set} in $\M$ whose distribution is defined
by $\bel$ in \eqref{eqbel}. Random sets will appear again in
Section \ref{SgeneralWB}.

The rigorous DS calculus laid out in Shafer \cite{shafer1976}, and
reformulated for statisticians in Dempster \cite{dempster2008}, makes
the DS analysis very attractive. A key element of the DS theory is
Dempster's rule of combination, which allows two (independent) pieces
of evidence, represented as belief functions on the same frame of
discernment, to be combined in a way that is similar to combining
probabilities via a product measure. While the intuition behind
Dempster's rule is quite simple, the general expression for the
combined belief function is rather complicated and is, therefore,
omitted; see Shafer \cite{shafer1976}, Chapter 3, or Yager and Liu
\cite{yagerliu2008}, Chapter~1, for the details. But in a statistical
context, the most important type of belief functions to be combined
with $\bel$ in \eqref{eqbel} are those that fix the value of either
the $X$ or $\Theta$ component---this type of combination is known as
\textit{conditioning}. It turns out that Dempster's rule of conditioning
is fairly simple; see Theorem 3.6 of Shafer~\cite{shafer1976}. Next we
outline the construction of these conditional belief functions,
handling the two distinct cases separately.

\subsubsection*{Condition on $\Theta$}
Here we combine the belief function \eqref{eqbel} with another based
on the information $\Theta= \theta$. Start with the trivial (constant)
set-valued mapping
\[
M_0(U) \equiv\{(X,\Theta)\dvtx  \Theta= \theta\}.
\]
This, together with the mapping $M$ in \eqref{eqfocal}, gives a
combined focal element
\[
M_0(U) \cap M(U) = \{(X,\theta)\dvtx  X = a(\theta,U)\},
\]
the $\theta$-cross section of $M(U)$, which we project down to the
$X$-margin to give
%
\begin{equation}
\label{eqdvtxdssampling}
M_\theta(U) = \{X\dvtx  X = a(\theta,U)\} \subset\X.
\end{equation}
Let $\A$ be a measurable subset of $\X$. It can be shown that the
conditional belief function $\bel_\theta$ can be obtained by applying
the same rule as in \eqref{eqbel} but with $M_\theta(U)$ in place of
$M(U)$. That is, the conditional belief function, given $\Theta= \theta
$, is given by
%
\begin{eqnarray}
\label{eqbeltheta}
\bel_\theta(\A) &=& \mu\{U\dvtx  M_\theta(U) \subseteq\A\}
\nonumber
\\[-8pt]
\\[-8pt]
\nonumber
&=& \mu\{U\dvtx  a(\theta
,U) \in\A\},
\end{eqnarray}
the push-forward measure defined by $\mu$ and the mapping $a(\theta
,\cdot)$, which is how the sampling distribution is defined. Therefore,
given $\Theta=\theta$, the conditional belief function $\bel_\theta
(\cdot)$ is just the sampling distribution $\Prob_\theta(\cdot)$.

\subsubsection*{Condition on $X$}
For given $X=x$, we proceed just as before; that is, start with the
trivial (constant) set-valued mapping
\[
M_0(U) \equiv\{(X,\Theta)\dvtx  X=x \}
\]
and combine this with $M(U)$ in \eqref{eqfocal} to obtain a new
posterior focal element
\[
M_0(U) \cap M(U) = \{(x,\Theta)\dvtx  x=a(\Theta,U)\},
\]
the $x$-cross section of $M(U)$, which we project down to the $\Theta$
margin to give
%
\begin{equation}
\label{eqdsposterior}
M_x(U) = \{\Theta\dvtx  x=a(\Theta,U)\} \subset\TT.
\end{equation}
Unlike the ``condition on $\Theta$'' case above, this posterior focal
element can, in general, be empty---a so-called \textit{conflict case}.
Dempster's rule of combination will effectively remove these conflict
cases by conditioning on the event that $M_X(U) \neq\varnothing$; see
Dempster \cite{dempster1967}. In this case, for an assertion, or
hypothesis, $\A\subset\TT$, the DS \textit{posterior belief function}
$\bel_x$ is defined as
%
\begin{equation}
\label{eqpostbel}
\bel_x(\A) = \frac{\mu\{U\dvtx  M_x(U) \subseteq\A\}}{\mu\{U\dvtx  M_x(U) \neq
\varnothing\}}.
\end{equation}

We now turn to some important properties of $\bel_x$. In Shafer's
axiomatic development, belief functions are \textit{nonadditive}, which implies
%
\begin{equation}
\label{eqnonadditive}
\bel_x(\A) + \bel_x(\A^c) \leq1\quad   \mbox{for all $\A$},
\end{equation}
with equality if and only if $\bel_x$ is an ordinary additive
probability. The intuition here is that evidence not in favor of $\A^c$
need not be in favor of $\A$. If we define the \textit{plausibility
function} as
%
\begin{equation}
\label{eqplaus}
\pl_x(\A) = 1-\bel_x(\A^c),
\end{equation}
then it is immediately clear from \eqref{eqnonadditive} that
\[
\bel_x(\A) \leq\pl_x(\A)\quad  \mbox{for all $\A$}.
\]
For this reason, $\bel_x(\A)$ and $\pl_x(\A)$ have often been called,
respectively, the \textit{lower} and \textit{upper probabilities} of $\A$
given $X = x$. In our statistical context, $\A$ plays the role of a
hypothesis about the unknown paramter $\Theta$ of interest. So for any
relevant assertion $\A$, the posterior belief and plausibility
functions $\bel_x(\A)$ and $\pl_x(\A)$ can be calculated, and
conclusions are reached based on the relative magnitudes of these quantities.

We have been writing ``$X=x$'' to emphasize that the posterior focal
elements and belief function are conditional on a fixed observed value
$x$ of $X$. But later we will consider sampling properties of the
posterior belief function, for fixed $\A$, as a function of the random
variable $X$, so, henceforth, we will write $M_X(U)$ for $M_x(U)$ in
\eqref{eqdsposterior}, and $\bel_X$ for $\bel_x$ in \eqref{eqpostbel}.

\begin{example}
\label{exnormalmean2}
Consider again the problem in Example \ref{exnormalmean} of making
inference on the unknown mean $\Theta$ of a Gaussian population
$N(\Theta,1)$ based on a single observation $X$. We can use the
a-equation $X = \Theta+ \Phi^{-1}(U)$, where $U \sim\mu= \Unif
(0,1)$. The focal elements $M(U)$ in \eqref{eqfocal} are the lines
\[
M(U) = \{(X,\Theta)\dvtx  X = \Theta+ \Phi^{-1}(U)\}.
\]
Given $X$, the focal elements $M_X(U) = \{X-\Phi^{-1}(U)\}$ in \eqref
{eqdsposterior} are singletons. Since $U \sim\Unif(0,1)$, the
posterior belief function
\[
\bel_X(\A) = \mu\{U\dvtx  X-\Phi^{-1}(U) \in\A\}
\]
is the probability that an $N(X,1)$ distributed random variable falls
in $\A$, which is the same as the objective Bayes and fiducial
posterior. Note also that this approach is different from that
suggested by Dempster~\cite{dempster1966} and described in detail in
Dempster \cite{dempster1969}.
\end{example}

\begin{figure}

\includegraphics{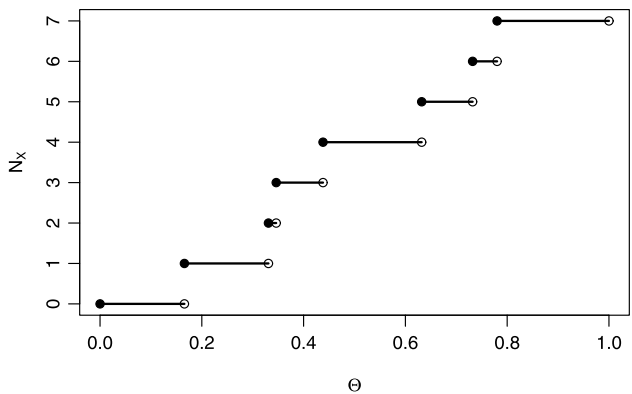}

\caption{A focal element $M(U)$ for the Bernoulli data problem in
Example \protect\ref{exbin}, with $n=7$. A posterior focal element is a
horizontal line segment, the $\Theta$-interval determined by fixing the
value of $N=N_{X}$.}
\label{figbinomfocal}
\end{figure}

\begin{example}
\label{exbin}
Suppose that the binary data $X = (X_1,\ldots,X_n)$ consists of
independent Bernoulli observations, and $\Theta\in[0,1]$ represents
the unknown probability of success. Dempster \cite{dempster1966}
considered the sampling model determined by the a-equation
%
\begin{equation}
X_i = I_{\{U_i \leq\theta\}}, \quad  i=1,\ldots,n,
\label{eqbernoullieqn}
\end{equation}
where $I_A$ denotes the indicator of the event $A$, and the auxiliary
variable $U = (U_1,\ldots,U_n)$ has pivotal measure $\mu= \Unif
([0,1]^n)$. The belief function will have generic focal elements
\[
M(U) = \bigl\{(X,\Theta)\dvtx  X_i = I_{\{U_i \leq\Theta\}} \mbox{ }  \forall
i=1,\ldots,n \bigr\}.
\]
This definition of the focal element is quite formal, but looking more
carefully at the a-equation \eqref{eqbernoullieqn} casts more light on
the relationships between $X_i$, $U_i$ and $\Theta$. Indeed, we know
that:
\begin{itemize}
\item if $X_i=1$, then $\Theta\geq U_i$, and
\item if $X_j = 0$, then $\Theta< U_j$.
\end{itemize}
Letting $N_X = \sum_{i=1}^n X_i$ be the number of successes in the $n$
Bernoulli trials, it is clear that exactly $N_X$ of the $U_i$'s are
smaller than $\Theta$, and the remaining $n-N_X$ are greater than
$\Theta$. There is nothing particularly important about the indices of
the $U_i$'s, so throwing out conflict cases reduces the problem from
the binary vector $X$ and uniform variates $U$ to the success count $N
= N_X$ and \textit{ordered} uniform variates; see Dempster \cite
{dempster1966} for a detailed argument. Let $U_{(i)}$ denote the
$i$th order statistic from $U_1,\ldots,U_n$, with $U_{(0)} :=
0$ and $U_{(n+1)} := 1$. Then the focal element $M(U)$ above reduces to
\begin{eqnarray*}
&M(U) = \bigl\{(N,\Theta)\dvtx  U_{(N)} \leq\Theta\leq
U_{(N+1)}\bigr\},&\\
&\hspace*{186pt}  U \in
[0,1]^n.&
\end{eqnarray*}
Figure \ref{figbinomfocal} gives a graphical representation of this
generic focal element. Now given $N$, the posterior belief function has
focal elements
%
\begin{eqnarray}
\label{eqbinomfocal}
&&M_N(U) = \bigl\{\Theta\dvtx  U_{(N)} \leq\Theta\leq U_{(N+1)} \bigr\},\\
\eqntext{ U \in[0,1]^n,}
\end{eqnarray}
which are intervals (the horizontal lines in Figure \ref
{figbinomfocal}) compared to the singletons in Example \ref
{exnormalmean2}. Consider the assertion $\A_\theta= \{\Theta\leq
\theta\}$ for $\theta\in[0,1]$. The posterior belief and plausibility
functions for $\A_\theta$ are given by
\begin{eqnarray*}
\bel_N(\A_\theta) & =& \mu\bigl\{U \in[0,1]^n\dvtx  U_{(N+1)} \leq\theta\bigr\}, \\
\pl_N(\A_\theta) & =& 1-\mu\bigl\{U \in[0,1]^n\dvtx  U_{(N)} > \theta\bigr\}.
\end{eqnarray*}
When $N$ is fixed, the marginal beta distributions of $U_{(N)}$ and
$U_{(N+1)}$ are available and $\bel_X(\A_\theta)$ and $\pl_X(\A_\theta
)$ can be readily calculated. Plots for the case of $n=12$ and observed
$N=7$ can be seen in Figure~\ref{figbinompqr} (Example \ref{exbinwb} in Section \ref{SSwb}).
\end{example}

Next are two important remarks about the conventional DS analysis just
described:
\begin{itemize}
\item The examples thus far have considered only ``dull'' assertions,
such as $\A= \{\Theta\leq\theta\}$, where conventional DS performs
fairly well. But for ``sharp'' assertions, such as $\A=\{\Theta= \theta
\}$, particularly in high-dimensional problems, conventional DS can be
too strong, resulting in plausibilities $\pl_X(\A) \approx0$ that are
of no practical use.
\item For fixed $\A$, $\bel_X(\A)$ has no built-in long-run frequency
properties as a function of $X$. Therefore, rules like ``reject $\A$ if
$\pl_X(\A) < 0.05$ or, equivalently, if $\bel_X(\A^c) \geq0.95$'' have
no guaranteed long-run error rates, so designing statistical\break \textit
{methodology} around conventional DS may be challenging.
\end{itemize}
It turns out that both of these problems can be taken care of by \textit
{shrinking} $\bel_X$ in \eqref{eqpostbel}. We do this in Section \ref
{SgeneralWB} by suitably weakening the conventional DS belief,
replacing the pivotal measure $\mu$ with a belief function.

\section{Inference with Weak Beliefs}
\label{SgeneralWB}

\subsection{Inferential Models}
\label{SSim}

The conventional DS analysis of the previous section achieves the lofty
goal of providing posterior probability-based inference without prior
specification, but the difficulties mentioned at the end of Section \ref
{SSds} have kept DS from breaking into the statistical mainstream. Our
basic premise is that these obstacles can be overcome by relaxing the
crucial ``continue to believe'' assumption. The concept of \textit
{inferential models} (IMs) will formalize this idea.

Let $\bel_X$ denote the posterior belief function \eqref{eqpostbel}
of the conventional DS analysis in Section \ref{SSds}, and let $\bel
^*$ be another belief function on the parameter space~$\TT$, possibly
depending on $X$. For any assertion $\A$ of interest, $\bel^*(\A)$ can
be calculated and, at least in principle, used to make inference on the
unknown~$\Theta$. We say that $\bel^*$ specifies an IM on $\TT$ if
%
\begin{equation}
\label{eqim}
\bel^*(\A) \leq\bel_X(\A)\quad   \mbox{for all $\A$}.
\end{equation}
Since $\bel^*$ has plausibility $\pl^*(\A) = 1-\bel^*(\A^c)$, it is
clear from \eqref{eqim} that $\pl^*(\A) \geq\pl_X(\A)$ for all $\A$.
Therefore, an IM can have meaningful nonzero plausibility even for
sharp assertions. Shrinking the belief function can be done by suitably
modifying the focal element mapping $M(\cdot)$ or the pivotal measure
$\mu$, but any other technique that generates a belief function bounded
by $\bel_X$ would also produce a valid IM.

$\bel_X$ itself specifies an IM, but is a very extreme case. At the
opposite extreme is the vacuous belief function with $\bel^*(\A) = 0$
for all $\A\neq2^\TT$. Clearly, neither of these IMs would be fully
satisfactory in general. The goal is to choose an IM that falls
somewhere in between these two extremes.

In the next subsection we use IMs to motivate the method of \textit{weak
beliefs}, due to Zhang and Liu \cite{zl2010}. That is, we apply their
WB method to construct a particular class of IMs and, in Section \ref
{SSmb}, we show how a particular IM can be chosen.

\subsection{Weak Beliefs}
\label{SSwbprinciple}

Section \ref{Sintro} described how the a-equation might be used for
data generation: fix $\Theta$, sample $U$ from the pivotal measure $\mu
$, and compute $X = a(\Theta,U)$. Now, for the inference problem,
suppose that the observed data~$X$ was, indeed, generated according to
this recipe, but the corresponding values of $\Theta$ and $U$ remain
hidden. Denote by $U^*$ the value of the unobserved auxiliary variable;
see \eqref{eqUstar}. The key point is that knowing $\Theta$ is
equivalent to knowing $U^*$; in other words, inference on $\Theta$ is
equivalent to \textit{predicting} the value of the unobserved $U^*$. Both
the fiducial and DS theories are based on this idea of shifting the
problem of inference on $\Theta$ to one of predicting $U^*$, although,
to our knowledge, neither method has been described in this way before.
The advantage of focusing on $U^*$ is that the a priori
distribution for $U^*$ is fully specified by the sampling model.

More formally, if the sampling model $\Prob_\Theta$ is specified by the
a-equation \eqref{eqaeqn}, then the following relation must hold \textit
{after} $X$ is observed:
%
\begin{equation}
\label{eqUstar}
X = a(\Theta,U^*),
\end{equation}
where $\Theta$ is unknown and $U^*$ is unobserved. We can ``solve''
this equation for $\Theta$ to get
%
\begin{equation}
\label{eqAeqn}
\Theta\in A(X,U^*),
\end{equation}
where $A(\cdot,\cdot)$ is a set-valued map. Intuitively, \eqref
{eqAeqn} identifies those parameter values which are consistent with
the observed $X$. For example, in the normal mean problem of
Example \ref{exnormalmean}, once $X$ has been observed, there is a
one-to-one relationship between the unknown mean $\Theta$ and the
unobserved $U^*$, that is, $\Theta= A(X,U^*) = \{X-\Phi^{-1}(U^*)\}$,
so, given $U^*$, one can immediately find $\Theta$. Therefore, if we
could predict $U^*$, then we could know $\Theta$ exactly. The crucial
``continue to believe'' assumption of fiducial and DS says that $U^*$
can be predicted by taking draws $U$ from the pivotal measure $\mu$. WB
weakens this assumption by replacing the draw $U \sim\mu$ with a set
$\S(U)$ containing $U$, which is equivalent to replacing $\mu$ with a
belief function.

Recall from Section \ref{SSds} that a measure and set-valued mapping
together define a belief function. Here we fix $\mu$ to be the pivotal
measure, and construct a belief function on $\U$ by choosing a
set-valued mapping $\S\dvtx  \U\to2^\U$ that satisfies $U \in\S(U)$. This
is not the same as the DS analysis described in Section \ref{SSds};
there the belief function was fully specified by the sampling model,
but here we must make a subjective choice of~$\S$. We call this pair
$(\mu,\S)$ a \textit{belief}, as it generates a belief function $\mu\S
^{-1}$ on $\U$. Intuitively, $(\mu,\S)$ determines how aggressive we
would like to be in predicting the unobserved $U^*$; more aggressive
means smaller $\S(U)$, and vice versa. We will call $\S(U)$, as a
function of $U \sim\mu$, a \textit{predictive random set} (PRS), and we
can think of the inference problem as trying to hit $U^*$ with the PRS
$\S(U)$.

The two extreme IMs---the DS posterior belief function $\bel_X$ in
\eqref{eqpostbel} and the vacuous belief function---are special cases
of this general framework; take $\S(U) = \{U\}$ for the former, and $\S
(U) = \U$ for the latter. So in this setting we see that the quality of
the IM is determined by how well the PRS $\S(U)$ can predict $U^*$.
With this new interpretation, we can explain the comment at the end of
Section \ref{SSds} about the quality of conventional DS for sharp
assertions in high-dimensional problems. Generally, high-dimensional
$\Theta$ goes hand-in-hand with\break high-dimensional $U$, and accurate
estimates of $\Theta$ require accurate prediction of $U^*$. But the
\textit{curse of dimensionality} states that, as the dimension increases,
so too does the probabilistic distance between $U^*$ and a random point
$U$ in $\U$. Consequently, the tiny (sharp) assertion $\A$ will rarely,
if ever, be hit by the focal elements $M_X(U)$.

In Section \ref{SSmb} we give a general WB framework, show how a
particular $\S$ can be chosen, and establish some desirable long-run
frequency properties of the weakened posterior belief function. But
first, in Section \ref{SSwb}, we develop WB inference for given $\S$
and give some illustrative examples.

\subsection{Belief Functions and WB}
\label{SSwb}

In this section we show how to incorporate WB into the DS analysis
described in Section \ref{SSds}. Suppose that a map $\S$ is given. The
case $\S(U) = \{U\}$ was taken care of in Section \ref{SSds}, so what
follows will be familiar. But this formal development of the WB
approach will highlight two interesting and important properties,
consequences of Dempster's conditioning operation.

Previously, we have taken the frame of discernment to be $\X\times\TT
$. Here we have additional uncertainty about $U^* \in\U$, so first we
will extend this to the larger frame $\X\times\TT\times\U$. The
belief function on $\U$ has focal elements
\[
\{U^* \in\U\dvtx  U^* \in\S(U)\},
\]
which correspond to cylinders in the larger frame, that is,
\[
\{(X,\Theta,U^*)\dvtx  U^* \in\S(U)\}.
\]
Likewise, extend the focal elements $M(U)$ in \eqref{eqfocal} to
cylinders in the larger frame with focal elements
\[
\{(X,\Theta,U^*)\dvtx  X = a(\Theta,U^*)\}.
\]
(The belief functions to which these extended focal elements correspond
are implicitly formed by combining the particular belief function with
the vacuous belief function on the opposite margin.) Combining these
extended focal elements, and simultaneously marginalizing over $\U$,
gives a new focal element on the original frame $\X\times\TT$, namely,
%
\begin{eqnarray}
\label{eqnewfocal}
\qquad M(U;\S) & =& \{(X,\Theta)\dvtx  X = a(\Theta,u), u \in\S(U)\}
\nonumber
\\[-8pt]
\\[-8pt]
\nonumber
& =& \bigcup\{M(u)\dvtx  u \in\S(U)\},
\end{eqnarray}
where $M(\cdot)$ is the focal mapping defined in \eqref{eqfocal}.
Immediately we see that the focal element $M(U;\S)$ in \eqref
{eqnewfocal} is an expanded version of $M(U)$ in \eqref{eqfocal}. The
measure $\mu$ and the mapping $M(U;\S)$ generate a new belief function
over $\X\times\TT$:
\[
\bel(\E;\S) = \mu\{U\dvtx  M(U;\S) \subseteq\E\}.
\]
Since $M(U) \subseteq M(U;\S)$ for all $U$, it is clear that $\bel(\E;\S
) \leq\bel(\E)$. The two DS conditioning operations will highlight the
importance of this point.

\subsubsection*{Condition on $\Theta$}
Conditioning on a fixed $\Theta= \theta$, the focal elements (as
subsets of $\X$) become
\begin{eqnarray*}
M_\theta(U;\S) & =& \{X\dvtx  X=a(\theta,u), u \in\S(U)\} \\
& = &\bigcup\{M_\theta(u)\dvtx  u \in\S(U)\}.
\end{eqnarray*}
This generates a new (predictive) belief function\break  $\bel_\theta(\cdot; \S
)$ that satisfies
\begin{eqnarray*}
\bel_\theta(\A;\S) & =& \mu\{U\dvtx  M_\theta(U;\S) \subseteq\A\} \\
& \leq&\mu\{U\dvtx  M_\theta(U) \subseteq\A\}\\
& =& \bel_\theta(\A) =\Prob
_\theta(\A).
\end{eqnarray*}
Therefore, in the WB framework, this conditional belief function need
not coincide with the sampling model as it does in the conventional DS
context. But the sampling model $\Prob_\theta(\cdot)$ is \textit
{compatible} with the belief function $\bel_\theta(\cdot;\S)$ in the
sense that
\[
\bel_\theta(\cdot;\S) \leq\Prob_\theta(\cdot) \leq\pl_\theta(\cdot;\S
).
\]
If we think about probability as a precise measure of uncertainty,
then, intuitively, when we weaken our measure of uncertainty about
$U^*$ by replacing $\mu$ with a belief function $\mu\S^{-1}$, we expect
a similar smearing of our uncertainty about the value of $X$ that will
be ultimately observed.

\subsubsection*{Condition on $X$}
Conditioning on the observed $X$, the focal elements (as subsets of $\TT
$) become
\begin{eqnarray*}
M_X(U;\S) & =& \{\Theta\dvtx  X=a(\Theta,u), u \in\S(U)\} \\
& =& \bigcup\{M_X(u)\dvtx  u \in\S(U)\}.
\end{eqnarray*}
Evidently, $M_X(U;\S)$ is just an expanded version of $M_X(U)$ in \eqref
{eqdsposterior}. But a larger focal element will be less likely to
fall completely within $\A$ or $\A^c$. Indeed, the larger $M_X(U;\S)$
generates a new posterior belief function $\bel_X(\cdot;\S)$ which satisfies
\begin{eqnarray} \label
{eqnewpostbel}
\qquad \bel_X(\A;\S) &=& \mu\{U\dvtx  M_X(U;\S) \subseteq\A\}
\nonumber
\\[-8pt]
\\[-8pt]
\nonumber
& \leq&\mu\{U\dvtx  M_X(U) \subseteq\A\} = \bel_X(\A).
\end{eqnarray}
Therefore, $\bel_X(\cdot;\S)$ is a bonafide IM according to \eqref{eqim}.

There are many possible maps $\S$ that could be used. In the next two
examples we utilize one relatively simple idea---using an
interval/rectangle\break $\S(U) = [A(U),B(U)]$ to predict $U^*$.

\begin{figure}

\includegraphics{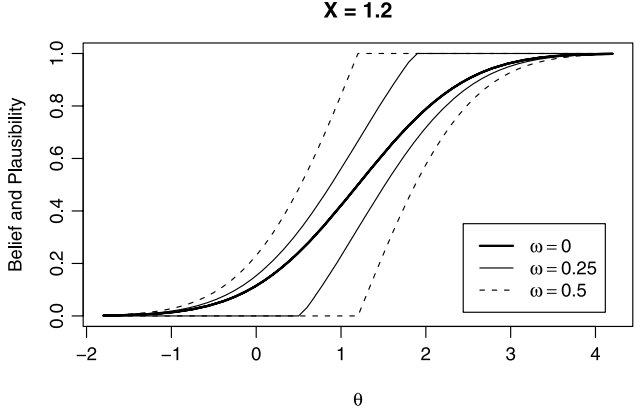}

\caption{Plots of belief and plausibility, as functions of $\theta$,
for assertions $\A_\theta= \{\Theta\leq\theta\}$ for $X = 1.2$ and
$\omega\in\{0,0.25,0.5\}$ in the normal mean problem in Example \textup{\protect\ref
{exnormalwb}}. The case $\omega= 0$ was considered in Example \textup{\protect\ref
{exnormalmean}}.}
\label{fignormalwb}
\end{figure}

\begin{example}
\label{exnormalwb}
Consider again the normal mean problem in Example \ref{exnormalmean}.
The posterior belief function was derived in Example \ref
{exnormalmean2} and shown to be the same as the objective Bayes
posterior. Here we consider a WB analysis where the set-valued mapping
$\S= \S_\omega$ is given by
%
\begin{eqnarray}
\label{eqSnormal}
&&\S(U) = [U-\omega U, U + \omega(1-U)], \\
\eqntext{  \omega\in[0,1].}
\end{eqnarray}
It is clear that the cases $\omega= 0$ and $\omega= 1$ correspond to
the conventional and vacuous beliefs, respectively. Here we will work
out the posterior belief function for $\omega\in(0,1)$ and compare
the result to that in Example~\ref{exnormalmean2}. Recall that the
posterior focal elements in Example~\ref{exnormalmean2} were
singletons $M_X(U) = \{\Theta\dvtx  \Theta= X-\Phi^{-1}(U)\}$. It is easy
to check that the weakened posterior focal elements are intervals of
the form
\begin{eqnarray*}
M_X(U;\S) & = &\bigcup\{M_X(u)\dvtx  u \in\S(U)\} \\
& =& \bigl[ X-\Phi^{-1}\bigl(U+\omega(1-U)\bigr),\\
&&\hspace*{33pt}{} X-\Phi^{-1}(U-\omega U) \bigr].
\end{eqnarray*}
Consider the sequence of assertions $\A_\theta= \{\Theta\leq\theta\}
$. We can derive analytical formulas for $\bel_X(\A_\theta)$ and $\pl
_X(\A_\theta)$ as functions of $\theta$:
%
\begin{eqnarray}
\label{eqnormbel}
\bel_X(\A_\theta;\S) & =& \biggl[ 1-\frac{\Phi(X-\theta)}{1-\omega}
\biggr]^+,
\nonumber
\\[-8pt]
\\[-8pt]
\nonumber
\pl_X(\A_\theta;\S) & =& 1-\biggl[ \frac{\Phi(X-\theta)-\omega}{1-\omega}
\biggr]^+,
\end{eqnarray}
where $x^+ = \max\{0,x\}$. Plots of these functions are shown in
Figure \ref{fignormalwb}, for $\omega\in\{0,0.25,0.5\}$, when $X =
1.2$ is observed. Here we see that as $\omega$ increases, the spread
between the belief and plausibility curves increases. Therefore, one
can interpret the parameter $\omega$ as a \textit{degree of weakening}.
\end{example}

\begin{figure}

\includegraphics{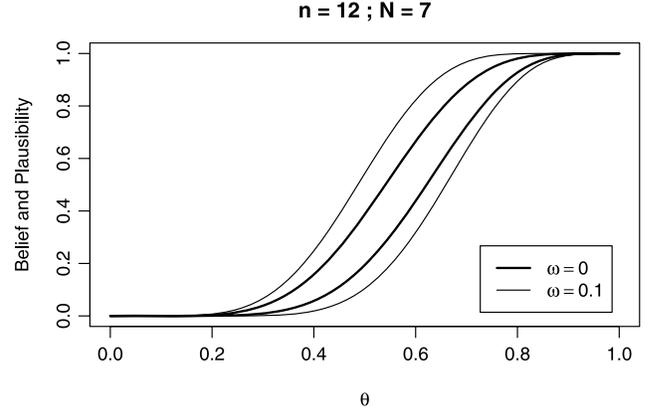}

\caption{Plots of belief and plausibility, as functions of $\theta$,
for assertions $\A_\theta= \{\Theta\leq\theta\}$ when $n=12$ and
$N=7$ and $\omega\in\{0,0.1\}$, in the Bernoulli success probability
problem in Example \textup{\protect\ref{exbinwb}}. The case $\omega= 0$ was considered
in Example \textup{\protect\ref{exbin}}.}
\label{figbinompqr}
\end{figure}

\begin{example}
\label{exbinwb}
Consider again the Bernoulli problem from Example \ref{exbin}. In this
setup, the auxiliary variable $U = (U_1,\ldots,U_n)$ in $\U= [0,1]^n$
is vector-valued. We apply a similar weakening principle as in
Example~\ref{exnormalwb}, where we use a rectangle to predict $U^*$.
That is, fix $\omega\in[0,1]$ and define $\S=\S_\omega$ as
\[
\S(U) = [A_1(U),B_1(U)] \times\cdots\times[A_n(U),B_n(U)]
\]
a Cartesian product of intervals like that in Example \ref
{exnormalwb}, where
\begin{eqnarray*}
A_i(U) & =& U_i - \omega U_i, \\
B_i(U) & =& U_i + \omega(1-U_i).
\end{eqnarray*}
Following the DS argument in Example \ref{exbin}, it is not difficult
to check that the (weakened) posterior focal elements are of the form
\begin{eqnarray*}
M_N(U;\S) &=& \bigl[ U_{(N)}-\omega U_{(N)}, U_{(N+1)}\\
&&\hspace*{26pt}{} + \omega
\bigl(1-U_{(N+1)}\bigr) \bigr],
\end{eqnarray*}
an expanded version of the focal element $M_X(U)$ in \eqref
{eqbinomfocal}. Computation of the belief and plausibility can still
be facilitated using the marginal beta distributions of $U_{(N)}$ and
$U_{(N+1)}$. For example, consider the sequence of assertions $\A
_\theta= \{\Theta\leq\theta\}$, $\theta\in[0,1]$. Plots of $\bel
_N(\A_\theta;\S)$ and $\pl_N(\A_\theta;\S)$, as functions of $\theta$,
are given in Figure \ref{figbinompqr} for $\omega= 0$ (which is the
conventional belief situation in Example \ref{exbin}) and $\omega=
0.1$, when $n=12$ and $N=7$. As expected, the distance between the
belief and plausibility curves is greater for the latter case. But this
naive construction of $\S$ is not the only approach; see Zhang and
Liu \cite{zl2010} for a more efficient alternative based on a
well-known relationship between the binomial and beta CDFs.
\end{example}

\subsection{The Method of Maximal Belief}
\label{SSmb}

The WB analysis for a given set-valued map $\S$ was described in
Section \ref{SSwb}. But how should one choose~$\S$ so that the
posterior belief function satisfies certain desirable properties?
Roughly speaking, the idea is to choose a map $\S$ with the
``smallest'' PRSs $\S(U)$ with the desired coverage probability.
Following Zhang and Liu \cite{zl2010}, we call this the method of \textit
{maximal belief} (MB).

Consider a general class of beliefs $\BB= (\mu,\SS)$, where $\mu$ is
the pivotal measure from Section \ref{Sintro}, and $\SS= \{\S_\omega\dvtx
\omega\in\Omega\}$ is a class of set-valued mappings indexed by
$\Omega$. Each $\S_\omega$ in $\SS$ maps points $u \in\U$ to subsets
$\S_\omega(u) \subset\U$ and, together with the pivotal measure $\mu$,
determines a belief function $\mu\S_\omega^{-1}$ on $\U$ and, in turn,
a posterior belief function $\bel_X(\cdot;\S_\omega)$ on $\TT$ as in
Section \ref{SSwb}. For a given class of beliefs, it remains to choose
a particular map $\S_\omega$ or, equivalently, an index $\omega\in
\Omega$, with the appropriate credibility and efficiency properties. To
this end, define
%
\begin{equation}
\label{eqqfunction}
Q_\omega(u) = \mu\{U\dvtx  \S_\omega(U) \not\ni u\}, \quad   u \in\U,
\end{equation}
which is the probability that the PRS $\S_\omega(U)$ misses the target
$u \in\U$. We want to choose $\S_\omega$ in such a way that the random
variable $Q_\omega(U^*)$, a function of $U^* \sim\mu$, is
stochastically small.

\begin{definition}
\label{defcredible}
A belief $(\mu,\S_\omega)$ is \textit{credible} at level $\alpha\in
(0,1)$ if
%
\begin{equation}
\label{eqcredible}
\phi_\alpha(\omega) := \mu\{U^*\dvtx  Q_\omega(U^*) \geq1-\alpha\} \leq
\alpha.
\end{equation}
\end{definition}

Note the similarity between credibility and the control of Type-I error
in the frequentist context of hypothesis testing. That is, if $\S_\omega
$ is credible at level $\alpha= 0.05$, then in a sequence of 100
similar inference problems, each having different $U^*$, we expect
$Q_\omega$---the probability that the PRS $\S_\omega$ misses its
target---to exceed 0.95 in no more than 5 of these cases. The analogy
with frequentist hypothesis testing is made here only to offer a way of
understanding credibility.

It is not immediately clear why this notion of credibility is
meaningful for the problem of inference on the unknown parameter $\Theta
$. The following theorem, an extension of Theorem 3.1 in Zhang and
Liu \cite{zl2010}, gives conditions under which $\bel_X(\cdot;\S)$ has
desirable long-run frequency properties in repeated $X$-sampling.

\begin{theorem}
\label{thmzlthm}
Suppose $(\mu,\S)$ is credible at level $\alpha\in(0,1)$, and that
$\mu\{U\dvtx  M_X(U;\S) \neq\varnothing\} = 1$. Then, for any assertion $\A
\subset\TT$, the posterior belief function $\bel_X(\A;\S)$ in \eqref
{eqnewpostbel}, as a function of $X$, satisfies
%
\begin{equation}
\label{eqdsfreq}
\qquad\Prob_\Theta\{\bel_X(\A;\S) \geq1-\alpha\} \leq\alpha, \quad  \Theta
\in\A^c.
\end{equation}
\end{theorem}

We can again make a connection to frequentist hypothesis testing, but
this time in terms of assertions/hypotheses $\A$ in the parameter
space. If we adopt the decision rule ``conclude $\Theta\notin\A$
if\break
$\pl_X(\A;\S) < 0.05$,'' then under the conditions of Theorem \ref
{thmzlthm} we have
\[
\Prob_\Theta\{ \pl_X(\A;\S) < 0.05\} \leq0.05, \quad  \Theta\in\A.
\]
That is, if $\A$ does contain the true $\Theta$, then we will
``reject'' $\A$ no more than 5\% of the time in repeated experiments,
which is analogous to Type-I error probabilities in the frequentist
testing domain. So the importance of Theorem \ref{thmzlthm} is that it
equates credibility of the belief $(\mu,\S)$ to long-run error rates of
belief/plausibility function-based decision rules. For example, the
belief $(\mu,\S_\omega)$ in \eqref{eqSnormal} is credible for $\omega
\in[0.5,1]$, so decision rules based on \eqref{eqnormbel} will have
controlled error rates in the sense of \eqref{eqdsfreq}. But remember
that belief functions are posterior quantities that contain
problem-specific evidence about the parameter of interest.

Credibility cannot be the only criterion, however since the belief, with $\S
(U) = \U$, is always credible at any level $\alpha\in(0,1)$. As an
analogy, a frequentist test with empty rejection region is certain to
control the Type-I error, but is practically useless; the idea is to
choose from those tests that control Type-I error one with the largest
rejection region. In the present context, we want to choose from those
$\alpha$-credible maps the one that generates the ``smallest'' PRSs. A
convenient way to quantify size of a PRS $\S_\omega(U)$, without using
the geometry of $\U$, is to consider its coverage probability
$1-Q_\omega$.

\begin{definition}
\label{defefficient}
$(\mu,\S_\omega)$ is as \textit{efficient} as $(\mu,\S_{\omega^\prime})$ if
\[
\phi_\alpha(\omega) \geq\phi_\alpha(\omega^\prime)\quad   \mbox{for
all $\alpha\in(0,1)$}.
\]
That is, the coverage probability $1-Q_\omega$ is (stochastically) no
larger than the coverage probability $1-Q_{\omega^\prime}$.
\end{definition}

Efficiency defines a partial ordering on those beliefs that are
credible at level $\alpha$. Then the level-$\alpha$ maximal belief
($\alpha$-MB) is, in some sense, the maximal $(\mu,\S_\omega)$ with
respect to this partial ordering. The basic idea is to choose, from
among those credible beliefs, one which is most efficient. Toward this,
let $\Omega_\alpha\subset\Omega$ index those maps $\S_\omega$ which
are credible at level $\alpha$.

\begin{definition}
\label{defmb}
For $\alpha\in(0,1)$, $\S_{\omega^*}$ defines an $\alpha$-MB if
%
\begin{equation}
\label{eqomegan0}
\phi_\alpha(\omega^*) = \sup_{\omega\in\Omega_\alpha} \phi_\alpha
(\omega).
\end{equation}
Such an $\omega^*$ will be denoted by $\omega(\alpha)$.
\end{definition}

By the definition of $\Omega_\alpha$, it is clear that the supremum on
the right-hand side of \eqref{eqomegan0} is bounded by $\alpha$.
Under fairly mild conditions on $\SS$, we show in Appendix \hyperref[SSexistence]{A.1} that there exists an $\omega^* \in\Omega_\alpha$ such that
%
\begin{equation}
\phi_\alpha(\omega^*) = \alpha, 
\label{eqomegan}
\end{equation}
so, consequently, $\omega^* = \omega(\alpha)$ specifies an $\alpha$-MB.
We will, henceforth, take \eqref{eqomegan} as our working definition
of MB. Uniqueness of a MB must be addressed case-by-case, but the
left-hand side of \eqref{eqomegan} often has a certain monotonicity
which can be used to show the solution is unique.

We now turn to the important point of computing the MB or,
equivalently, the solution $\omega(\alpha)$ of the equation~\eqref
{eqomegan}. For this purpose, we recommend the use of a \textit
{stochastic approximation} (SA) algorithm, due to Robbins and
Monro \cite{robbinsmonro}. Kushner and Yin \cite{kushner} give a
detailed theoretical account of SA, and Martin and Ghosh \cite{martinghosh} give an
overview and some recent statistical applications.

Putting all the components together, we now summarize the four basic
steps of a MB analysis:
\begin{enumerate}
\item Form a class $\BB= (\mu,\SS)$ of candidate beliefs, the choice
of which may depend on (a) the assertions of interest, (b) the nature
of your personal uncertainty, and/or (c) intuition and
geometric/computational simplicity.
\item Choose the desired credibility level $\alpha$.
\item Employ a stochastic approximation algorithm to find an $\alpha
$-MB as determined by the solution of \eqref{eqomegan}.
\item Compute the posterior belief and plausibility functions via Monte
Carlo integration by simulating the PRSs $\S_{\omega(\alpha)}(U)$.
\end{enumerate}

In Sections \ref{Stesting} and \ref{Snonparametrics} we will describe
several specific classes of beliefs and the corresponding PRSs. These
examples certainly will not exhaust all of the possibilities; they do,
however, shed light on the considerations to be taken into account when
constructing a class $\BB$ of beliefs.

\begin{figure*}[b]

\includegraphics{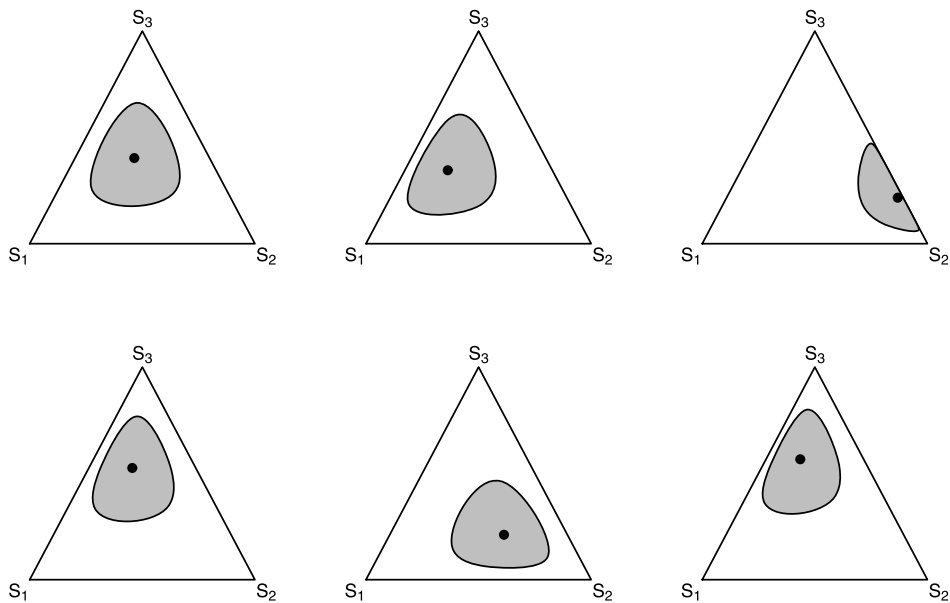}

\caption{Six realizations of $R$-cross sections of the PRS $\S_\omega
(R,P)$ in (\protect\ref{eqauxSRS1}) in the case of $n=3$. Here $\PP_2$ is the
triangular region in the Barycentric coordinate system.}
\label{figKLPRS}
\end{figure*}

\section{High-Dimensional Testing}
\label{Stesting}

A major focus of current statistical research is very-high-dimensional
inference and, in particular, multiple testing. This is partly due to
new scientific technologies, such as DNA microarrays and medical
imaging devices, that give experimenters access to enormous amounts of
data. A typical problem is to make inference on an unknown $\Theta\in
\R^n$ based on an observed $X \sim N_n(\Theta,I_n)$; for example,
testing $H_{0i}\dvtx  \Theta_i = 0$ for each $i=1,\ldots,n$. See Zhang and
Liu \cite{zl2010} for a maximal belief solution of this
many-normal-means problem. Below we consider a related
problem---testing homogeneity of a Poisson process.

Suppose we monitor a system over a pre-specified interval of time, say,
$[0,\tau]$. During that period of time, we observe $n$ events/arrivals
at times $0 = \tau_0 < \tau_1 < \tau_2 < \cdots< \tau_n$, where the
$(n+1)$st event, taking place at $\tau_{n+1} > \tau$, is
unobserved. Assume an exponential model for the inter-arrival times
$X_i = \tau_i-\tau_{i-1}$, $i=1,\ldots,n$, that is,
%
\begin{equation}
\label{eqprocess1}
X_i \sim\Exp(\Theta_i),\quad    i=1,\ldots,n,
\end{equation}
where the $X_i$'s are independent and the exponential rates $\Theta
_1,\ldots,\Theta_n > 0$ are unknown. A question of interest is whether
the underlying process is homogeneous, that is, whether the rates
$\Theta_1,\ldots,\Theta_n$ have a common value. This question, or
hypothesis, corresponds to the assertion
%
\begin{eqnarray}
\label{eqhomogeneityassertion}
\A&=& \{\mbox{the process is homogeneous}\}
\nonumber
\\[-8pt]
\\[-8pt]
\nonumber
&=& \{\Theta_1 = \Theta_2 =
\cdots= \Theta_n\}.
\end{eqnarray}

Let $(X,\Theta)$ be the real-world quantities of interest, where $X =
(X_1,\ldots,X_n)$, $\Theta= (\Theta_1,\ldots,\Theta_n)$ and $\X= \TT
= (0,\infty)^n$. Define the auxiliary variable $U = (R,P)$, where $R >
0$ and $P = (P_1,\ldots,P_n)$ is in the $(n-1)$-dimensional probability
simplex $\PP_{n-1} \subset\R^n$, defined as
\[
 \PP_{n-1} = \Biggl\{(p_1,\ldots,p_n) \in[0,1]^n\dvtx  \sum
_{i=1}^n p_i = 1 \Biggr\}.
\]
The variables $R$ and $P$ are functions of the data $X_1,\ldots,X_n$
and the parameters $\Theta_1,\ldots,\Theta_n$. The a-equation $X =
a(\Theta,U)$, in this case, is given by $X_i = R P_i / \Theta_i$, where
%
\begin{eqnarray}
\label{eqauxmapping1}
R &=& \sum_{j=1}^n \Theta_j X_j \quad  \mbox{and}
\nonumber
\\[-8pt]
\\[-8pt]
\nonumber
P_i &= &\frac{\Theta
_i X_i}{\sum_{j=1}^n \Theta_j X_j},\quad    i=1,\ldots,n.
\end{eqnarray}
To complete the specification of the sampling model, we must choose the
pivotal measure $\mu$ for the auxillary variable $U=(R,P)$. Given the
nature of these variates, a natural choice is the product measure
%
\begin{equation}
\label{eqauxmeasure1}
\mu= \operatorname{Gamma}(n,1) \times\Unif(\PP_{n-1}).
\end{equation}
The measure $\mu$ in \eqref{eqauxmeasure1} is, indeed, consistent
with the exponential model \eqref{eqprocess1}. To see this, note that
$\Unif(\PP_{n-1})$ is equivalent to the Dirichlet distribution $\operatorname
{Dir}(1_n)$, where $1_n$ is an $n$-vector of unity. Then, conditional
on $(\Theta_1,\ldots,\Theta_n)$, it follows from standard properties of
the Dirichlet distribution that $\Theta_1 X_1,\ldots,\Theta_nX_n$ are
i.i.d. $\Exp(1)$, which is equivalent to \eqref{eqprocess1}.

We now proceed with the WB analysis. Step 1 is to define the class of
mappings $\SS$ for prediction of the unobserved auxiliary variables
$U^*=(R^*,P^*)$. To expand a random draw $U = (R,P) \sim\mu$ to a
random set, consider the class of maps $\SS= \{\S_\omega\dvtx  \omega\in
[0,\infty]\}$ defined as
%
\begin{eqnarray}
\label{eqauxSRS1}
\S_\omega(U) &=& \{ (r,p) \in[0,\infty)\times\PP_{n-1}\dvtx
\nonumber
\\[-8pt]
\\[-8pt]
\nonumber
&&\hspace*{50pt}{}  K(P,p) \leq
\omega\},
\end{eqnarray}
where $K(P,p)$ is the Kullback--Leibler (KL) divergence
%
\begin{equation}
\label{eqKL}
\hspace*{26pt} K(P,p) = \sum_{i=1}^n P_i \log(P_i/p_i),\quad    p,P \in\PP_{n-1}.
\end{equation}

Several comments on the choice of PRSs \eqref{eqauxSRS1} are in
order. First, notice that $\S_\omega(U)$ does not constrain the value
of $R$, that is, $\S_\omega(U)$ is just a cylinder in $[0,\infty)
\times\PP_{n-1}$ defined by the $P$-component of $U$. Second, the use
of the KL divergence in \eqref{eqauxSRS1} is motivated by the
correspondence between $\PP_{n-1}$ and the set of all probability
measures on $\{1,2,\ldots,n\}$. The KL divergence is a convenient tool
for defining neighborhoods in $\PP_{n-1}$. Figure \ref{figKLPRS}
shows cross-sections of several random sets $\S_\omega(U)$ in the case
of $n=3$.

\begin{figure*}[b]

\includegraphics{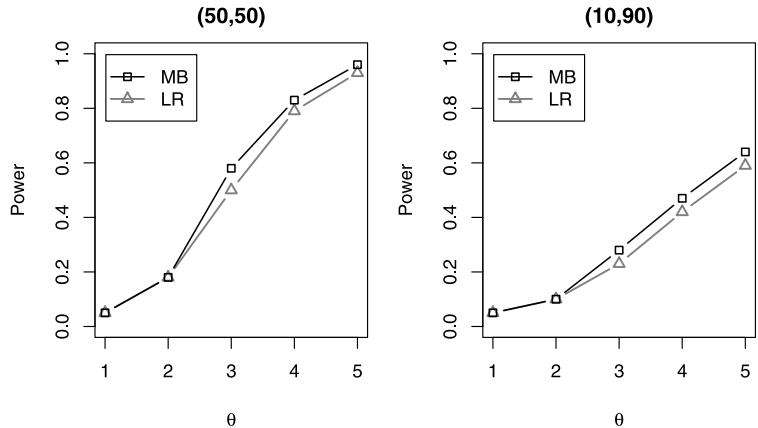}

\caption{Power of the MB and LR tests of homogeneity in Example \protect\ref
{exMBLR}, where $\theta$ is the ratio of the rate for the last $n_2$
observations to the rate of the first $n_1$ observations. Left:
$(n_1,n_2) = (50,50)$. Right: $(n_1,n_2) = (10,90)$.}
\label{figMBLR}
\end{figure*}

After choosing a credibility level $\alpha\in(0,1)$, we are on to
Step 3 of the analysis: finding an $\alpha$-MB. As in Section \ref
{SgeneralWB}, define
\[
Q_\omega(r,p) = \mu\{(R,P)\dvtx  \S_\omega(R,P) \not\ni(r,p) \},
\]
and, finally, choose $\omega= \omega(\alpha)$ to solve the equation
\[
\mu\{(R^*,P^*)\dvtx  Q_\omega(R^*,P^*) \geq1-\alpha\} = \alpha.
\]
This calculation requires stochastic approximation.

For Step 4,\vspace*{1pt} first define the mapping $\widehat{P}\dvtx  \TT\to\PP_{n-1}$\break
by the component-wise formula $\widehat{P}_i(\Theta) = \Theta_i X_i /\break
\sum_j \Theta_j X_j$, $i=1,\ldots,n$. For inference on $\Theta= (\Theta
_1,\break \ldots,\Theta_n)$, a posterior focal element is of the form
\[
M_X\bigl(R,P;\S_{\omega(\alpha)}\bigr) = \{ \Theta\dvtx  K(P,\widehat{P}(\Theta)) \leq
\omega(\alpha) \}.
\]
For the homogeneity assertion $\A$ in \eqref{eqhomogeneityassertion}
the posterior belief function is zero, but the plausibility is given by
\begin{eqnarray*}
&&\pl_X\bigl(\A;\S_{\omega(\alpha)}\bigr) \\
&&\quad= 1-\mu\{(R,P)\dvtx  K(P,\widehat{P}(1_n)) >
\omega(\alpha) \},
\end{eqnarray*}
where $\widehat{P}_i(1_n) = X_i/\sum_j X_j$. Since $\widehat{P}(1_n)$
is known and $P \sim\Unif(\PP_{n-1})$ is easy to simulate, once $\omega
(\alpha)$ is available, the plausibility can be readily calculated
using Monte Carlo.

In order to assess the performance of the MB method above in testing
homogeneity, we will compare it with the typical likelihood ratio (LR)
test. Let $\ell(\Theta)$ be the likelihood function under the general
model \eqref{eqprocess1}. Then the LR test statistic for $H_0\dvtx  \Theta
_1 = \cdots= \Theta_n$ is given by
\begin{eqnarray*}
L_0 &=& \frac{\sup\{\ell(\Theta)\dvtx  \Theta\in H_0\}}{\sup\{\ell(\Theta)\dvtx
\Theta\in H_0 \cup H_0^c\}}\\
& =& \biggl[ \frac{ ( \prod_{i=1}^n X_i
)^{1/n}}{\overline{X}} \biggr]^n,
\end{eqnarray*}
a power of the ratio of the geometric and arithmetic means. If $\widehat
{P}$ is as defined before, then a little algebra shows that
\[
L = -\log L_0 = nK(u_n,\widehat{P}(1_n)),
\]
where $u_n$ is the $n$-vector $n^{-1} 1_n$ which corresponds to the
uniform distribution on $\{1,2,\ldots,n\}$. Note that this problem is
invariant under the group of scale transformations, so the null
distribution of $\widehat{P}(1_n)$ and, hence~$L$, is independent of
the common value of the rates $\Theta_1,\ldots,\Theta_n$. In fact,
under the homogeneity assertion \eqref{eqhomogeneityassertion},
$\widehat{P}(1_n) \sim\Unif(\PP_{n-1})$.

\begin{example}
\label{exMBLR}
To compare the MB and LR tests of homogeneity described above, we
performed a simulation. Take $n = n_1+n_2=100$, $n_1$ of the rates
$\Theta_1,\ldots,\Theta_n$ to be 1 and $n_2$ of the rates to be $\theta
$, for various values of $\theta$. For each of 1000 simulated data
sets, the plausibility for $\A$ in \eqref{eqhomogeneityassertion}. To
perform the hypothesis test using $q$, we choose a nominal 5\% level
and say ``reject the homogeneity hypothesis if plausibility~$< 0.05$.''
The power of the two tests are summarized in Figure \ref{figMBLR},
where we see that the MB test is noticeably better than the LR test.
The MB test also controls the frequentist Type-I error at 0.05. But
note that, unlike the LR test, the MB test is based on a meaningful
data-dependent measure of the amount of evidence supporting the
homogeneity assertion.
\end{example}

\section{Nonparametrics}
\label{Snonparametrics}

A fundamental problem in nonparametric inference is the so-called \textit
{one-sample test}. Specifically, assume that $X_1,\ldots,X_n$ are i.i.d.
observations from a distribution on $\R$ with CDF $F$ in a class $\FF$
of CDFs; the goal is to test $H_0\dvtx  F \in\FF_0$ where $\FF_0 \subset\FF
$ is given. One application is a test for normality, that is, where $\FF
_0 = \{N(\theta,\sigma^2)$ for some $\theta$ and $\sigma^2$\}.
This is an important problem, since many popular methods in applied
statistics, such as regression and analysis of variance, often require
an approximate normal distribution of the data, of residuals, etc.

We restrict attention to the simple one-sample testing problem, where
$\FF_0 = \{F_0\} \subset\FF$ is a singleton. Our starting point is the
a-equation
%
\begin{equation}
\label{eqfishernp}
X_i = F^{-1}(U_i), \quad   F \in\FF,   i=1,\ldots,n,
\end{equation}
where $U_1,\ldots,U_n$ are i.i.d. $\Unif(0,1)$. Since $F$ is monotonically
increasing, it is sufficient to consider the ordered data $X_{(1)} \leq
X_{(2)} \leq\cdots\leq X_{(n)}$, the corresponding ordered auxiliary
variables $\Utilde= (U_{(1)},\ldots,U_{(n)})$, and pivotal measure\vspace*{1pt} $\mu
$ determined by the distribution of $\Utilde$.

In this section we present a slightly different form of WB analysis
based on \textit{hierarchical} PRSs. In hierarchical Bayesian analysis, a
random prior is taken to add an additional layer of flexibility. The
intuition here is similar, but we defer the discussion and technical
details to Appendix \ref{SShprs}.

For predicting $\Utilde^*$, we consider a class of beliefs indexed by
$\Omega= [0,\infty]$, whose PRSs are small $n$-boxes inside the unit
$n$-box $[0,1]^n$. Start with a fixed set-valued mapping that takes
ordered $n$-vectors $\tilde u \in[0,1]^n$, points $z \in(0.5,1)$, and
forms the intervals $[A_i(z),B_i(z)$], where
%
\begin{eqnarray}
\label{eqAiBi}
\qquad A_i(z) & =& \operatorname{qBeta}(p_i-zp_i \mid i, n+1-i),
\nonumber
\\[-8pt]
\\[-8pt]
\nonumber
B_i(z) & =& \operatorname{qBeta}\bigl(p_i+z(1-p_i) \mid i,n+1-i\bigr)
\end{eqnarray}
and $p_i = \operatorname{pBeta}(u_{(i)} \mid i,n-i+1)$. Here $\operatorname{pBeta}$ and
$\operatorname{qBeta}$ denote CDF and inverse CDF of the Beta distribution,
respectively. Then the mapping $\S(\tilde u,z)$ is just the Cartesian
product of these $n$ intervals; cf. Example~\ref{exbinwb}. Now sample
$\Utilde$ and $Z$ from a suitable distribution depending on $\omega$:
\begin{itemize}
\item Take a draw $\Utilde$ of $n$ ordered $\Unif(0,1)$ variables.
\item Take $V \sim\Bet(\omega,1)$ and set $Z = \frac{1}{2}(1+V)$.
\end{itemize}
The result is a random set $\S(\Utilde,Z) \in2^\U$. We call this
approach ``hierarchical'' because one could first sample $Z=z$ from the
transformed beta distribution indexed by $\omega$, fix the map $\S(\cdot
,z)$, and then sample $\Utilde$.\vspace*{1pt}

For a draw $(\Utilde,Z)$, the posterior focal elements for~$F$ look like
\begin{eqnarray*}
&&M_X(\Utilde,Z;\S) = \bigl\{F\dvtx  A_i(Z) \leq F\bigl(X_{(i)}\bigr) \leq B_i(Z), \\
&&\hspace*{130pt}\qquad  \forall
  i=1,\ldots,n\bigr\}.
\end{eqnarray*}
Details of the credibility of in a more general context are given in
Appendix \hyperref[SShprs]{A.2}. Stochastic approximation is used, as in
Section \ref{Stesting}, to optimize the choice of $\omega$. The MB
method uses the posterior focal elements above, with optimal $\omega$,
to compute the posterior belief and plausibility functions for the
assertion $\A= \{F = F_0\}$ of interest.

\begin{figure*}

\includegraphics{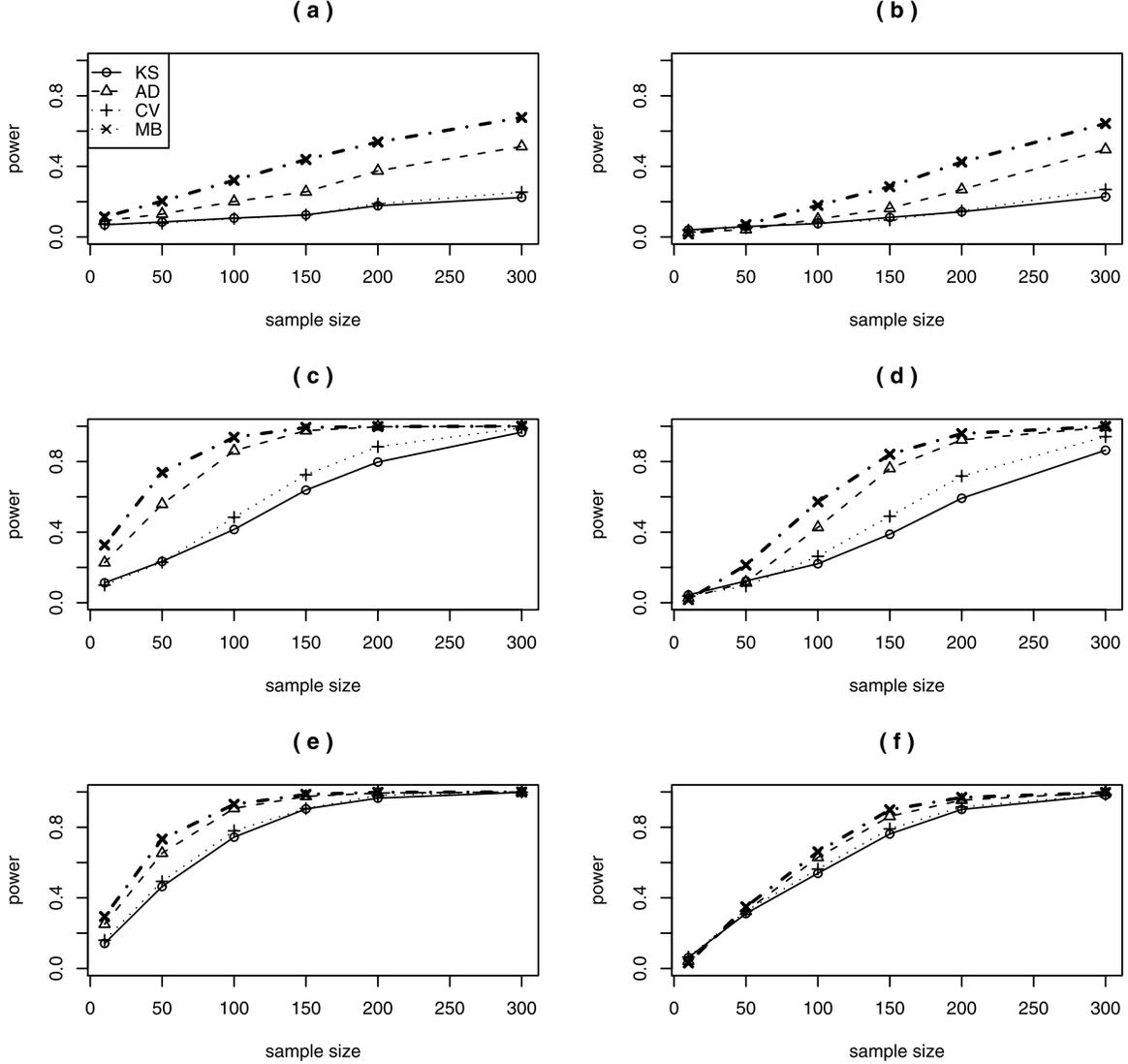}

\caption{Power comparison for the one-sample tests in Example \protect\ref
{exnp} at level $\alpha= 0.05$ for various values of
$n$. The six alternatives are
\textup{(a)}~$\operatorname{Beta}(0.8,0.8)$;
\textup{(b)}~$\operatorname{Beta}(1.3,1.3)$;
\textup{(c)}~$\operatorname{Beta}(0.6,0.6)$;
\textup{(d)}~$\operatorname{Beta}(1.6,1.6)$;
\textup{(e)}~$\operatorname{Beta}(0.6,0.8)$;
\textup{(f)}~$\operatorname{Beta}(1.3,1.6)$.
}
\label{fignp}
\end{figure*}

\begin{example}
\label{exnp}
To illustrate the performance of the MB method, we present a small
simulation study. We take $F_0$ to be the CDF of a $\Unif(0,1)$
distribution. Samples $X_1,\ldots,X_n$, for various sample sizes~$n$,
are taken from several nonuniform distributions and the power of MB,
along with some of the classical tests, is computed. We have chosen our
nonuniform alternatives to be $\operatorname{Beta}(\beta_1,\beta_2)$ for
various values of $(\beta_1,\beta_2)$. For the MB test, we use the
decision rule ``reject $H_0$ if plausibility $< 0.05$.'' Figure \ref
{fignp} shows the power of the level $\alpha=0.05$ Kolmogorov--Smirnov
(KS), Anderson--Darling (AD), Cram\'{e}r--von Mises (CV) and MB tests,
as functions of the sample size $n$ for six pairs of $(\beta_1,\beta
_2)$. From the plots we see that the MB test outperforms the three
classical tests in terms of power in all cases, in particular, when $n$
is relatively small and the alternative is symmetric and ``close'' to
the null [i.e., when $(\beta_1,\beta_2) \approx(1,1)$]. Here, as in
Example \ref{exMBLR}, the MB test also controls the Type-I error at
level $\alpha= 0.05$.
\end{example}

\section{Discussion}
\label{Sdiscuss}

In this paper we have considered an modification of the DS theory in which
some desired frequency properties can be realized while, at the same
time, the essential components of DS inference, such as ``don't know,''
remain intact. The WB method was justified within a more general
framework of inferential models, where posterior probability-based
inference with frequentist properties is the primary goal. In two
high-dimensional hypothesis testing problems, the MB method
performs quite well compared to popular frequentist methods in terms of
power---more work is needed to fully understand this relationship
between WB/MB hypothesis testing and frequentist power. Also, the
detail in which these examples were presented should shed light on how
MB can be applied in practice.

One potential criticism of the WB method is the lack of uniqueness of
the a-equations and PRS mappings~$\SS$. At this stage, there are no
optimality results justifying any particular choices. Our approach thus
far has been to consider relatively simple and intuitive ways of
constructing PRSs, but further research is needed to define these
optimality criteria and to design PRSs that satisfy these criteria.

In addition to the applications shown above, preliminary results of WB
methods in other statistical problems are quite promising. We hope that
this work on WBs will inspire both applied and theoretical
statisticians to take a another look at what DS has to offer.

\begin{appendix}

\section*{Appendix: Technical results}

\subsection{Existence of a MB}
\label{SSexistence}

Consider a class $\SS= \{\S_\omega\dvtx  \omega\in\Omega\}$ of set-valued
mappings. Assume that the index set $\Omega$ is a complete metric
space. Each $\S_\omega$, together with the pivotal measure $\mu$,
define a belief function $\mu\S_\omega^{-1}$ on $\U$. Here we show that
there is a $\omega= \omega(\alpha)$ that solves the equation~\eqref{eqomegan}.
To this end, we make the following assumptions:
\begin{enumerate}[A1.]
\item[A1.] Both the \textit{conventional} and \textit{vacuous} beliefs are
encoded in $\SS$.
\item[A2.] If $\omega_n \to\omega$, then $\S_{\omega_n}(u) \to\S
_\omega(u)$ for each $u \in\U$.
\end{enumerate}
Condition A1 is to make sure that $\BB$ is suitably rich, while A2
imposes a sort of continuity on the sets \mbox{$\S_\omega\in\SS$}.

\begin{proposition}
Under assumptions \textup{A1--A2},\break  there exists a solution $\omega(\alpha
)$ to \eqref{eqomegan} for any $\alpha\in(0,1)$.
\end{proposition}

\begin{pf}
For notational simplicity, we write $Q(\omega,\break u)$ for $Q_\omega(u)$. We
start by showing $Q(\omega,u)$ is continuous in $\omega$. Choose
$\omega\in\Omega$ and a sequence $\omega_n \to\omega$. Then under A2
\begin{eqnarray*}
Q(\omega_n,u) &=& \int I_{\{\S_{\omega_n}(v) \not\ni u\}}  \,d\mu(v) \\
&\to&
\int I_{\{\S_\omega(v) \not\ni u \}} \, d\mu(v) = Q(\omega,u)
\end{eqnarray*}
by the dominated convergence theorem (DCT). Since $\omega_n \to\omega$
was arbitrary and $\Omega$ is a metric space, it follows that $Q(\cdot
,u)$ is continuous on $\Omega$.

Write $\phi(\omega)$ for $\phi_\alpha(\omega)$ in \eqref{eqcredible};
we will now show that $\phi(\cdot)$ is continuous. Again choose $\omega
\in\Omega$ and a sequence $\omega_n \to\omega$. Define $J_\omega(u) =
I_{\{Q(\omega,u) \geq1-\alpha\}}$, so that $\phi(\omega) = \int
J_\omega(u)  \,d\mu(u)$. Since
\[
|\phi(\omega_n) - \phi(\omega)| \leq\int| J_{\omega_n}(u)-J_\omega
(u)|\, d\mu(u)
\]
and the integrand on the right-hand side is bounded by~2, it follows,
again follows by the DCT, that $\phi(\omega_n) \to\phi(\omega)$ and,
hence, that $\phi(\cdot)$ is continuous on $\Omega$. But A1 implies
that $\phi(\cdot)$ takes values 0 and 1 on $\Omega$ so by the
intermediate value theorem, for any $\alpha\in(0,1)$, there exists a
solution $\omega= \omega(\alpha)$ to the equation $\phi(\omega) =
\alpha$.
\end{pf}

\subsection{Hierarchical PRSs}
\label{SShprs}

In Section \ref{Snonparametrics} we considered a WB analysis with
hierarchical PRSs. The purpose of this generalization is to provide a
more flexible choice of random sets for predicting the unobserved
$U^*$. Here we give a theoretical justification along the lines in
Section \ref{SSmb}.

Let $\omega\in\Omega$ index a family of probability measures $\lambda
_\omega$ on a space $\ZZ$,\vspace*{1pt} and suppose $\S(\cdot,\cdot)$ is a fixed
set-valued mapping $\U\times\ZZ \to2^\U$, assumed to satisfy $U \in
\S(U,Z)$ for all $Z$. A hierarchical PRS is defined by first taking $Z
\sim\lambda_\omega$ and then choosing the map $\S_Z(\cdot) = \S(\cdot
,Z)$ defined on $\U$. This amounts to a product pivotal measure $\mu
\times\lambda_\omega$. Toward credibility of $(\mu\times\lambda
_\omega,\S)$, define the noncoverage probability
\begin{eqnarray*}
\Qbar_\omega(u) &=& (\mu\times\lambda_\omega)\{(U,Z)\dvtx  \S(U,Z) \not\ni
u\}\\
&= &\int Q_z(u)  \,d\lambda_\omega(z),
\end{eqnarray*}
a mixture of the noncoverage probabilities in \eqref{eqqfunction}.
Then we have the following, more general, definition of credibility.

\begin{definition}
\label{defnewcred}
$(\mu,\S_\omega)$ is credible at level $\alpha$ if
\[
\phibar_\alpha(\omega) := \mu\{U^*\dvtx  \Qbar_\omega(U^*) \geq1-\alpha\}
\leq\alpha.
\]
\end{definition}

Beliefs which are credible in the sense of Definition~\ref
{defcredible} are also credible according to Definition \ref
{defnewcred}---take $\lambda_\omega$ to be a point mass at $\omega$.
It is also clear that if $(\mu,\S_z)$ is credible in the sense of
Definition \ref{defcredible} for all $z \in\ZZ$, then $(\mu\times
\lambda_\omega,\S)$ will also be credible. Next we generalize
Theorem \ref{thmzlthm} to handle the case of hierarchical PRSs.

\begin{theorem}
\label{thmhprs}
Suppose that $(\mu\times\lambda_\omega,\S)$ is credible at level
$\alpha$ in the sense of Definition\textup{ \ref{defnewcred}}, and that $(\mu
\times\lambda_\omega)\{(U,Z)\dvtx  M_X(U, Z;\S) \neq\varnothing\} = 1$. Then
for any assertion $\A\subset\TT$, the belief function $\bel_X(\A; \S)
= (\mu\times\lambda_\omega)\S^{-1}(\A)$ satisfies
\[
\Prob_\Theta\{\bel_X(\A;\S) \geq1-\alpha\} \leq\alpha,\quad   \Theta
\in\A^c.
\]
\end{theorem}

\begin{pf}
Start by fixing $Z = z$, and write $\S_z(\cdot) = \S(\cdot,z)$. For
$\Theta\in\A^c$, monotonicity of the belief function gives
\begin{eqnarray*}
\bel_X(\A;\S_z) &\leq&\bel_X(\{\Theta\}^c;\S_z) \\
&= &\mu\{U\dvtx  M_X(U;\S_z)
\not\ni\Theta\}.
\end{eqnarray*}
When $\Theta$ is the true value, the event $M_X(U; \S_z) \not
\ni\Theta$ is equivalent to $\S_z(U) \not\ni U^*$; consequently,
\[
\bel_X(\A;\S_z) \leq\mu\{U\dvtx \S_z(U) \not\ni U^*\} = Q_z(U^*).
\]
For the hierarchical PRS, the belief function satisfies
\begin{eqnarray*}
\bel_X(\A;\S) & =& (\mu\times\lambda_\omega)\{(U,Z)\dvtx  M_X(U,Z;\S)
\subseteq\A\} \\
& =& \int\mu\{U\dvtx  M_X(U;\S_z) \subseteq\A\} \, d\lambda_\omega(z) \\
& = &\int\bel_X(\A;\S_z)  \,d\lambda_\omega(z) \\
& \leq&\int Q_z(U^*)  \,d\lambda_\omega(z) \\
& = &\Qbar_\omega(U^*).
\end{eqnarray*}
The claim now follows from credibility of the belief $(\mu\times
\lambda_\omega,\S)$.
\end{pf}
\end{appendix}

\section*{Acknowledgments}

The authors would like to thank Professor A. P. Dempster for sharing
his insight, and also the Editor, Associate Editor and three referees
for helpful suggestions and criticisms.

\end{document}